\documentclass[aps,pre,twocolumn,groupedaddress,showpacs,showkeys,floatfix]{revtex4-1}
\usepackage{wrapfig}
\usepackage{graphicx}
\usepackage{colordvi}
\usepackage{color}
\usepackage{amsmath}
\usepackage{amssymb}
 \usepackage{comment}
\usepackage{natbib}

\usepackage[small, bf]{caption}

\newcommand{\ve}[1]{\mbox{\boldmath $#1$}}

\begin{document}
\title{Theoretical Analysis of Discreteness-Induced Transition \\in Autocatalytic Reaction Dynamics}
 

\author{Nen Saito}
\email[]{saito@complex.c.u-tokyo.ac.jp, Tel.: +81-3-5454-6732}
\affiliation{
Graduate School of Arts and Sciences, The University of Tokyo, 3-8-1 Komaba, Meguro-ku, Tokyo 153-8902, Japan}
\author{Kunihiko Kaneko}
\email[]{kaneko@complex.c.u-tokyo.ac.jp, Tel.: +81-3-5454-6746} 
\affiliation{
Graduate School of Arts and Sciences,
The University of Tokyo, 3-8-1 Komaba, Meguro-ku, Tokyo 153-8902, Japan}
\date{\today}
\begin{abstract}
Transitions in the qualitative behavior of chemical reaction dynamics with a decrease in molecule number have attracted much attention. Here, a method based on a Markov process with a tridiagonal transition matrix is applied to the analysis of this transition in reaction dynamics. The transition to bistability due to the small-number effect and the mean switching time between the bistable states are analytically calculated in agreement with numerical simulations. In addition, a novel transition involving the reversal of the chemical reaction flow is found in the model under an external flow, and also in a three-component model. The generality of this transition and its correspondence to biological phenomena are also discussed.
\end{abstract}
\maketitle

\section{Introduction}
Temporal changes in chemical concentrations are often analyzed by using the rate equation of the reaction kinetics
in which a set of deterministic ordinary differential equations is adopted.
Fluctuation around the average change in the concentration is neglected by assuming that the total number of molecules is
sufficiently large; however, stochasticity in the reaction due to fluctuation in the number of reactants does exist, and it
is non-negligible, especially when the number of molecules is small. Such stochasticity can introduce a qualitative change
in the behavior of the reaction dynamics.
The fluctuation around the average behavior is typically analyzed using the linear noise approximation (LNA)~\cite{van1992stochastic}, which is represented by the Langevin equation with additive noise or the corresponding Fokker--Planck equation derived from van Kampen's system-size expansion~\cite{van1992stochastic,kubo1973fluctuation}.
Recently, several theoretical studies have reported ``small-number effects'' or ``discreteness-induced 
transitions" that lead to a qualitative deviation from the behavior expected by the LNA, due 
to the small number of components; these effects have been reported in catalytic reaction 
dynamics~\cite{togashi2001transitions,ohkubo2008transition,biancalani2012noise,togashi2003alteration,awazu2009self,
awazu2007discreteness,samoilov2005stochastic,remondini2013analysis,kobayashi2011connection}, reaction-diffusion systems~\cite{shnerb2000importance,togashi2004molecular, 
butler2011fluctuation}, gene regulatory circuits~\cite{ma2012small}, and 
ecology~\cite{mckane2005predator,biancalani2014noise}.

One remarkable example of a discreteness-induced transition, the emergence of multi-stability, was reported by Togashi and 
Kaneko~\cite{togashi2001transitions}. In their investigation of a catalytic chemical reaction system composed of four 
chemical species, it was revealed that when the total number of molecules is small, the system exhibits temporal switching 
between quasi-steady states in which only two chemical species are abundant and the others are extinct. The model was then 
simplified into two components and has been analyzed~\cite{ohkubo2008transition,biancalani2012noise,biancalani2014noise}, but
the small-number regime requires further mathematical analysis.

In this paper, we apply a method based on a Markov process with a tridiagonal transition matrix in order to analyze the small-number effect in chemical reactions, as well as investigate a novel type of discreteness-induced transition in chemical current.

The organization of this paper is as follows. 
In Sec. II, we demonstrate the validity of the method in an analysis of the two-component Togashi--Kaneko (2TK) model, and then, in Sec. III, introduce a 3-component model that exhibits a novel discreteness-induced transition: the current of the chemical reaction reverses when the total number of molecules becomes small. By applying both the method proposed here and the Fokker--Planck equation, the transition caused by a decrease in the number of molecules is explained with quantitative agreement with numerical simulations.
In Sec. IV, we show that the 2-component Togashi-Kaneko model under external flow also exhibits the reversal of chemical current.
In Sec. V, we give concluding remarks.
\section{Analysis of two-component Togashi--Kaneko (2TK) model}
\begin{figure}
\includegraphics[width=8cm]{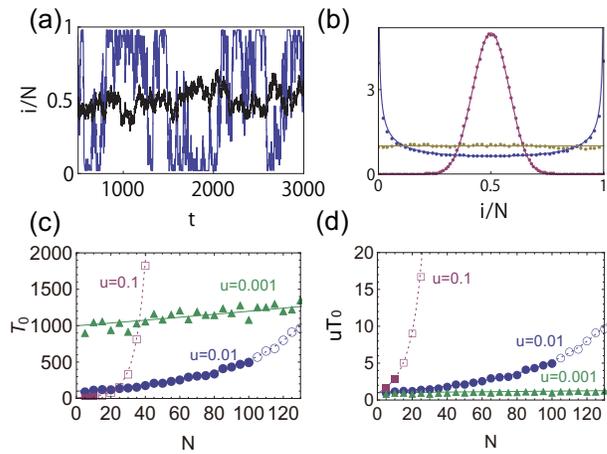}
\caption{Behavior of the 2TK model. (a) Time series of the concentration of A molecules for $N=50$ (blue line) and $N=2000$ (black line). $u$ is set as $u=0.01.$ The fixed point predicted by the rate equation is $i/N=0.5$.
(b) Steady state distributions of $i/N$. The unimodal distribution, the uniform distribution and the bimodal distribution indicate the steady state distribution for $N=2000$, $100$, and $50$, respectively. Dots represent simulation results and lines indicate analytical results obtained from the previous studies~\cite{ohkubo2008transition,biancalani2014noise}.
(c) Dependence of the switching time $T_{0}$ on $N$ and (d) that of the scaled switching time $uT_{0}$ on $N$. Top to bottom: $u=0.1$, $0.01$, and $0.001$. 
The symbols represent simulated results; the filled symbols denote those for $N \le 1/u$, where switching behavior appears.
For the region with $N>1/u$, $T_{0}$ increases drastically as $N$ increases, because $i/N$ stays at around $0.5$ for the majority of the time (as shown in (a)) and, accordingly, a significantly longer time period is required for $i/N=0$ or $1$ to be realized.}
\label{fig:2stateTKs}
\end{figure}

The 2TK model~\cite{ohkubo2008transition,biancalani2014noise} consists of the following four chemical reactions involving the 
two chemicals A and B: 
\begin{eqnarray}\label{eq:TK2}
&A+B \xrightarrow[k]{} 2A &, \ \ \  A+B \xrightarrow[k]{} 2B, \ \ \  A \overset{v}{\underset{u}{\leftrightarrows}} B,
\end{eqnarray}	
where $k$, $u$, and $v$ are the rate constants of the reactions. 
Note that the total number of molecules, denoted by $N$, is conserved in this model.
Without loss of generality, we can set $k=1$ 
by rescaling the time scale and set $u/k \to u$ and $v/k \to v$. 
By denoting the number of molecules of A and B as $i$ and $N-i$, respectively, the model can then
be defined in the state space $i=0, 1, \dots N$.
The transition rates from state $i$ to $i+1$ and from $i$ to $i-1$ are given by
$\lambda_{i}=i(N-i)/N+v(N-i)$ and $\mu_{i}=i(N-i)/N +ui$,
where the volume of the system is set to be $N$.
Examples of the time series of $i/N$ for $u=v=0.01$, obtained numerically using the Gillespie 
algorithm~\cite{gillespie1977exact}, are shown in Fig.~\ref{fig:2stateTKs}(a). Although the result for large $N$ ($N=2000$), 
shown by the black line, converges to the fixed-point concentration $i/N = v/(u+v)$ predicted by the rate equation, switching 
behavior between $i/N=0$ and $i/N=1$ emerges for small $N$ ($N=50$), as shown by the blue line. 
Correspondingly, the steady state distribution of $i/N$ shows a transition from a unimodal distribution ($N=2000$) to a bimodal distribution ($N=50$), as shown in Fig.~\ref{fig:2stateTKs}-(b).
The emergence of this switching behavior is an example of a discreteness-induced transition~\cite{togashi2001transitions} or noise-induced 
bistability~\cite{biancalani2014noise}.
Previous studies~\cite{biancalani2012noise,biancalani2014noise} give an analytical calculation of the steady state distribution and the critical value of $N$ for the 
appearance of switching behavior to be $N_{c}\simeq 1/u$ for the case $v=u$.
It is especially noteworthy that this model is similar to the two-allele Moran model with mutation in population 
genetics; The Moran process describes the neutral evolution of a population under a fixed population size; an 
individual agent is randomly replaced by another in each generation. The 2TK model is almost equivalent to the two-allele Moran model with bidirectional mutation. The difference lies in the use of continuous and discrete time. Indeed, with a decrease in $N$, the Moran model is known to show a transition from a state with two coexisting alleles to a state with alternate fixation of only one of the two alleles~\cite{ewens2004mathematical,gillespie2010population}. This corresponds to the switching behavior of the 2TK model. 
In addition, both the critical value of $N$ and the steady-state distribution calculated using the Moran model coincide with 
those of the 2TK model~\cite{ohkubo2008transition,biancalani2014noise}.

Although an analysis based on the Fokker--Planck equation is often adopted for chemical reaction systems, 
in general, it is not applicable when $N$ is small, the case in which we are interested.
When the system in question is described by a single-variable Markov process with a tridiagonal transition matrix, 
one can analytically obtain basic characteristic quantities.
To illustrate this point, we estimate the mean switching time from state $i = 0$ to state $i = N$. 
Similar treatments have been adopted in population genetics~\cite{ewens2004mathematical}, as well as  in physics~\cite{van1992stochastic}.
Setting $i=0$ as the initial condition and $i=N$ as the final absorbing state, the chemical reaction system in Eq.~(\ref{eq:TK2}) is described by the master equation: 
$\dot{P_{i}}=\lambda_{i-1}P_{i-1}+\mu_{i+1}P_{i+1}-(\lambda_{i}+\mu_{i})P_{i}$ with the boundary conditions $\mu_{0}=\mu_{N}=0$ and $\lambda_{N}=0$, where $P_{i}$ represents the probability of the state $i$ at time $t$.
We consider the occupancy time $t_{ij}$, that is, the mean time spent in state $j$ starting from state $i$ before absorption, which is defined by 
$t_{ij} \equiv \int_{0}^{\infty}dt P_{i}$. By integrating the master equation with respect to time $t$ from $t=0$ to $\infty$, we obtain
\begin{equation}\label{eq:tij}
t_{ij}=\mu_{i}t_{i-1,j}+(1-\mu_{i}-\lambda_{i})t_{i,j}+\lambda_{i}t_{i+1,j}+\delta_{ij},
\end{equation}
where $\delta_{ij}$ is the Kronecker delta.
Noting that the boundary conditions are $\mu_{N}=0$ and $t_{N,j}=0$ for $j=0, \dots N$,
the time $t_{0,j}$ is expressed as
 \begin{equation}\label{eq:t0j}
 t_{0,j}=\frac{1}{\lambda_{j}}\sum_{l=j}^{N-1} \prod_{k=j+1}^{l} \frac{\mu_{k}}{\lambda_{k}},
 \end{equation}
where $\prod _{k=j+1}^{j}(\mu_{k}/\lambda_{k})=1$. 
The switching time is thus given by $T_{0}=\sum_{i=0}^{N-1}t_{0,i}$. Figure~\ref{fig:2stateTKs}(c) and (d) show the estimated $T_{0}$ and those scaled by $u$, respectively.  Both agree well with the results obtained from numerical simulations.
Note that Biancalani et al.~\cite{biancalani2014noise} calculated the mean switching time using the Fokker--Planck equation for large $N$ and estimated the corresponding value for small $N$ in a heuristic manner. Our treatment, in contrast, gives a single expression, Eq.~(\ref{eq:t0j}), that shows remarkable agreement for all $N$.
It should be emphasized here that the expressions in Eqs.~(\ref{eq:tij}) and (\ref{eq:t0j}) are not limited to this specific model but are generally applicable to any $\lambda_{i}$ and $\mu_{i}$.

\begin{figure}[t]
\includegraphics[width=8cm]{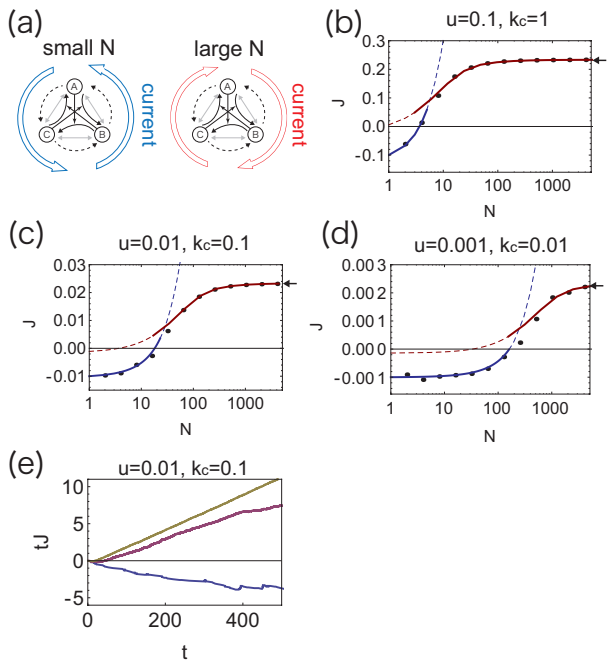}
\caption{Behavior of the 3-component model. (a) Schematic representation of the model. Gray double-headed arrows represent the autocatalytic reactions, while the black arrows represent the catalytic reactions (e.g., $A+C \to B+C$). Dashed arrows indicate the one-body reactions. The net reaction current $J$ flows clockwise for large $N$ (right) and counterclockwise for small $N$ (left). (b)--(d) Dependence of $J$ on $N$ for $(u,k_{c})=(0.1,1), (0.01, 0.1)$, and $(0.001, 0.01)$. Simulated results are shown by the black circles; theoretical estimates of $J$ from Eq.~(\ref{eq:Jfp}) and Eq.~(\ref{eq:Jtij}) are shown by the thick red and blue lines, respectively; and dashed lines indicate regions in which the conditions for the theoretical estimates are invalid. The $J$ values of the fixed-point solutions of the rate equation are indicated by the arrows.
(e) Time series of the total current that flows clockwise in the 3-component model for $N=1,000$, $100$, and $10$ (from top to bottom), with parameters $u=0.01$ and $k_{c}=0.1$. The net current $J$ is given by the slope of this time series. 
Larger fluctuation in the slope is observed for smaller $N$, which indicates that the net current is reversed in a stochastic manner.}
\label{fig:TK3}
\end{figure}

\section{three-component model and reversal of chemical current}
Provided a given chemical reaction system is described by a single-variable
Markov process with a tridiagonal transition matrix, we can rigorously apply the present analytical formulation.
The present formulation is also applicable to a multi-variable Markov process as an approximation, if one can properly project the multi-variable case onto a single variable Markov process.
Now, we demonstrate that this approximation is indeed valid when the number of molecules is small and discreteness-induced transition occurs.
As examples, we introduce here a 3-component model with autocatalytic reaction motifs in this section and also a 2TK model under external flow (i.e., without conservation of $N$) in the next section. 
In both cases, the chemical current exists due to breaking of the detailed balance and its reversal occurs as a novel discreteness-induced transition.
On one hand, these two examples demonstrate the general validity of our method while, on the other hand, they exhibit novel transitions by the small-number effect, beyond the emergence of the bistability discussed thus far.

The 3-component model, illustrated in Fig.~\ref{fig:TK3}(a), involves the following reactions of the three chemical species A, B and C in a container with a volume $N$:
\begin{eqnarray}\label{eq:TK3}
A+B \xrightarrow[1]{} 2A/2B, \ \ &B+C \xrightarrow[1]{} 2B/2C, \ \ &A + C \xrightarrow[1]{} 2A/2C,  \nonumber \\ 
A \xrightarrow[u]{} C, &C \xrightarrow[u]{} B,  & B \xrightarrow[u]{} A, \\
A+C \xrightarrow[k_{c}]{} B+C, \ \ &B+A \xrightarrow[k_{c}]{} C+A,  \ \ &C+B \xrightarrow[k_{c}]{} A+B. \nonumber
\end{eqnarray}	
Let $a$, $b$, and $c$ be the number of molecules of $A$, $B$, and $C$, respectively. The transition rates $T_{ (a', b', c' | a,b,c)}$ from state $(a,b,c)$ to state $(a',b',c')$ are then given as
\begin{eqnarray}\label{eq:tranTK3}
T_{(a+1,b-1,c | a, b, c)} &=u b+\frac{ab}{N}, \ &T_{(a-1, b+1, c | a , b, c)} =k_{c} \frac{ac}{N}+\frac{ab}{N}, \nonumber \\
T_{(a+1, b, c-1| a, b, c)} &=k_{c} \frac{bc}{N}+\frac{ac}{N}, \ &T_{( a-1, b, c+1| a,  b, c)} =u a+\frac{ac}{N}, \\
T_{(a,b+1,c-1 | a, b, c)} &=u c+\frac{bc}{N}, \ &T_{(a, b-1, c+1 | a, b, c)} =k_{c} \frac{ab}{N}+\frac{bc}{N},   \nonumber 
\end{eqnarray}
Note that $N$ is conserved.
The rate equation of this model is given by
\begin{eqnarray}\label{eq:3TKrate}
\dot{x}=u(y-x)+k_{c} (yz-xz),\\ \nonumber
\dot{y}=u(z-y)+k_{c} (xz-xy),\\ \nonumber
\dot{z}=u(x-z)+k_{c} (xy-yz),
\end{eqnarray}
where $x$, $y$, and $z$ are the concentrations of A, B, and C, respectively.
Since the detailed balance condition is not satisfied, a cyclic molecular current can exist, and we define the net reaction current $J$ as
$J=J_{a \to b}+J_{b \to c} +J_{c \to a}$. Here, 
$J_{a \to b}$ is given by 
\begin{equation}\label{eq:Jab}
J_{a \to b} = \sum_{a, b} P_{st} (a,b,c) \left[  T_{(a-1, b+1, c | a,b,c)}-  T_{(a+1, b-1, c | a,b,c)} \right], 
\end{equation}
where $c=N-a-b$, and $P_{st} (a,b,c)$ is the probability of state $(a, b, c)$ in the steady state;
$J_{b \to c}$ and $J_{c \to a}$ are given in a similar fashion. 
The net current $J$ for the rate equation in Eq.~(\ref{eq:3TKrate}) is calculated as $J=k_{c}/3-u$, where $P_{st} (a,b,c)$ is given by the $\delta$-function at the fixed-point solution $x=y=z=1/3$.

Although the rate equation in Eq.~(\ref{eq:3TKrate}) predicts a stationary clockwise current $J>0$ for $k_{c} >3u$, the numerical simulation gives a counterclockwise current $J<0$ when the total number of molecules $N$ is small. As shown in
Fig.~\ref{fig:TK3}(b)-(e), the net current $J$ decreases with decreasing $N$ and becomes negative.
Hence, the small-number effect leads to a reversal of the net current of the chemical reaction. 
Fluctuation in the total current, shown in Fig.~\ref{fig:TK3}-(e), indicates that the net current becomes negative in a stochastic manner, which is analogous to the fact that bimodal steady state distribution induced by the small-number effect in 2TK model appears in a stochastic manner.
Note that, although $J=0$ is satisfied if the detailed balance condition is satisfied, the opposite is not true; the detailed balance is not necessarily satisfied at the point $J=0$, because $J$ is defined by the average value for all $(a,b,c)$.
We analyze this transition by using both the Fokker--Planck equation derived from the master equation for large $N$ and the proposed method for small $N$.
The master equation of this model is expressed as
\begin{equation}
\dot{P}(x,y,t)= \sum_{x',y'}  \left[  T_{(xy|x'y')}P(x',y',t)  - T_{(x'y'|xy)}P(x,y,t)  \right], 
\end{equation}
where $T_{(xy|x'y')}$ is obtained from Eq.~(\ref{eq:tranTK3}) by letting $x=a/N$, $y=b/N$, and $z=c/N=1-x-y$. Using a Kramers--Moyal expansion to second order, the Fokker--Planck equation can be obtained from the master equation as 
\begin{eqnarray} \label{eq:FP}
&\dot{P}(x,y,t)&=\left[ -\frac{\partial}{\partial x}M_{x}-\frac{\partial}{\partial y}M_{y} \right. \\ \nonumber
&&\left. +   \frac{1}{2}\frac{\partial^{2}}{\partial x^{2}}M_{xx}+\frac{1}{2}\frac{\partial^{2}}{\partial y^{2}}M_{yy}+\frac{\partial^{2}}{\partial x\partial y}M_{xy}\right] P(x,y,t),
\end{eqnarray}
where 
\begin{eqnarray}
M_{x}&=&(y-x)\{ k_{c}(1-x-y)+u\}, \nonumber \\
M_{y}&=&(1-x-2y)\{ k_{c}x+u\}, \nonumber \\
M_{xx}&=&N^{-1}\left[ (k_{c}(1-x-y) +u)(x+y)+2x(1-x)\right],  \nonumber \\
M_{yy}&=&N^{-1}\left[ (k_{c}x +u)(1-x)+2y(1-y)\right],  \nonumber \\
M_{xy}&=&N^{-1}\left[-u y-k_{c}x(1-x-y)-2xy\right].
\end{eqnarray}
Here, the stationary solution of Eq.~(\ref{eq:FP}) cannot be obtained analytically, since the transition rates do not satisfy the detailed balance condition.
As an analytical estimate, we approximate the stationary solution of Eq.~(\ref{eq:FP}) by a Gaussian distribution of mean $(m_{x},m_{y})$ and variance-covariance matrix $\Sigma$, as $P_{0}(x,y)\propto \exp \left[ -\ve{v} \Sigma^{-1}\ve{v} /2\right]$, where $\ve{v}=(x-m_{x},y-m_{y})$. Under this approximation, 
$m_{x}$, $m_{y}$, the variances $V_{x}$ and $V_{y}$, and the covariance $V_{xy}$ are obtained as $m_{x}=m_{y}=1/3, V_{x}=V_{y}=-2V_{xy}=2(2+k_{c}+3u)/9(2+k_{c} +k_{c}N+3uN )$.
Note that this Gaussian approximation is invalid when $V_{x}$ or $V_{y}$ are large because the estimated probability distribution $P_{st}(x,y)$ then extends beyond the domain $\Omega( x, y | 0 \le x+y \le 1, \ x \ge 0, y \ge 0)$.
Thus, $V_{x}\le m_{x}^{2}$ is a necessary condition for the validity of the Gaussian approximation.

The net current estimated from the Gaussian approximation is expressed as 
\begin{eqnarray}\label{eq:Jfp}
J &=&\int_{\Omega} dx dy \left[  k_{c}\{ x(1-x)+y(1-y)-xy \} -u \right] P_{0}(x,y) \nonumber \\ 
&-&u \left( 1-\int_{\Omega} dx dy  P_{0}(x,y) \right).
\end{eqnarray}
The second term in this equation is a result of replacing the probability mass of $P_{0}(x,y)$ outside of $\Omega$ with a probability of states with $(x,y)=(1,0)$, $(0,1)$, or $(0,0)$.
Hence a negative current is predicted, especially when a large amount probability mass is outside of the $\Omega$ domain due to a large $V_{x}$ and $V_{y}$.
As shown in Fig.~\ref{fig:TK3}(b), (c), and (d), the current $J$ calculated using Eq.~(\ref{eq:Jfp}) agrees rather well with the numerical results, as long as $V_{x}\le m_{x}^{2}$ holds. As is seen in Fig. \ref{fig:TK3} (b)-(d), the invalid region of Eq.~(\ref{eq:Jfp}) also predicts a negative current. This is attributed to the second term of Eq.~(\ref{eq:Jfp}); the probability mass of $P_{0} (x,y)$ outside $\Omega$ is regarded as the state $(x,y)=(1,0)$, $(0,1)$, or $(0,0)$, and thus the current can be negative.

For small $N$, the variance is much larger, and so the Fokker--Planck equation is no longer valid.
 The proposed method, however, is applicable. In contrast to the case with the 2TK model, the 3-component model has two degrees of freedom, and thus the relationship in Eq.~(\ref{eq:tij}) cannot be applied straightforwardly.
Nevertheless, we can apply Eq.~(\ref{eq:tij}) by focusing only on a relatively short time scale within which the process can be approximated as a single-variable Markov process.
\begin{figure}
\includegraphics[width=8.5cm]{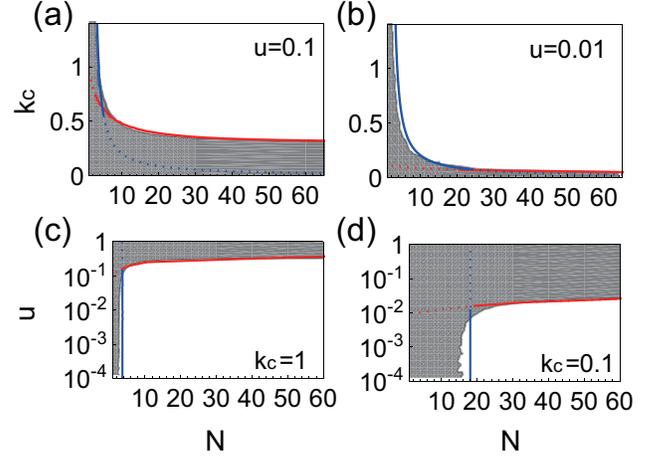}
\caption{Phase diagrams of the 3-component model in the (a),(b) $N$-$k_{c}$ plane and (c),(d) $N$-$u$ plane. 
White and shaded regions represent positive and negative $J$ (clockwise and counterclockwise current), respectively, as determined by the simulations. Thick red and blue lines indicate the theoretical results from Eq.~(\ref{eq:Jfp}) and  Eq.~(\ref{eq:Jtij}), respectively, while the dotted lines indicate the results for which the validity condition for the theoretical estimate is not satisfied.}
\label{fig:pds}
\end{figure}
In the present model, autocatalytic reactions (e.g., $A + C \to 2A/2C$) tend to push the system into a state in which only a single type of molecule dominates, while the other reactions promote the coexistence of the three molecular species.
Therefore, the model can be regarded as a molecule (e.g., A or C) extinction process if the autocatalytic reactions dominate.

Let us assume that only A molecules exist in the initial condition and that the system transits from $(a,b,c)=(N,0,0)$ to $(a,b,c)=(N-1,0,1)$ by the reaction $A \to C$ at $t=0$. 
If $u$ is sufficiently small, one can assume that one-body reactions (e.g., $A \to C$) do not occur within the time interval $1/Nu$ but that a molecule of A or C becomes extinct within the interval.
These assumptions allow us to approximate the behavior of the model by a single-variable Markov process with the transition probabilities $\lambda_{i}=i(N-i)/N$ and $\mu_{i}=(1+k_{c})i(N-i)/N$,
where $i$ now represents the number of A molecules. In these equations, both $i=0$ and $i=N$ are absorbing states.
During this transition process, any B molecules that are generated are counted as C molecules.
The time $t_{N-1,j}$ spent in state $j$ before the absorption starting from the initial state $i=N-1$ is calculated using Eq.~(\ref{eq:tij}) with the boundary condition $t_{0,j}=t_{N,j}=0$; 
\begin{equation}
t_{N-1,j}=\frac{\left( \prod_{k=j+1}^{N-1}  \frac{\mu_{k}}{\lambda_{k}}\right) }{\lambda_{j}} 
\frac{\sum_{l=1}^{j} \left( \prod_{k=1}^{l-1}  \frac{\mu_{k}}{\lambda_{k}}\right)}{\sum_{l=1}^{N} \left( \prod_{k=1}^{l-1}  \frac{\mu_{k}}{\lambda_{k}}\right)}=\frac{1}{\alpha \lambda_{j}}\frac{1-\alpha^{-j}}{1-\alpha^{-N}},
\end{equation}
for $k_{c} \neq 0$, where $ \alpha=1+k_{c}$ and $t_{N-1,j}=(N-j)^{-1}$ for $k_{c}=0$.
The total numbers of molecules that flow clockwise and counterclockwise within the interval $1/Nu$ are then calculated as $I_{+}=\sum_{j=1}^{N-1}(1+k_{c})j(N-j)t_{N-1,j}/N$ and $I_{-}=1+\sum_{j=1}^{N-1}j(N-j)t_{N-1,j}/N$, respectively.
The net current $J$ is evaluated from $u(I_{+}-I_{-})$ as
\begin{equation}\label{eq:Jtij}
J=-u+\sum_{j=1}^{N-1}u k_{c} \frac{j(N-j)}{N}t_{N-1,j}.
\end{equation}
This estimate is valid only when the time until absorption, $T_{N-1}=\sum_{j=1}^{N-1}t_{N-1,j}$, is smaller than $1/Nu$. The current $J$, shown by the blue lines in Fig.~\ref{fig:TK3}(b), (c), and (d), agrees rather well with the numerical results when $T_{N-1}<1/Nu$ holds. 

Figure~\ref{fig:pds} shows phase diagrams of the regime for clockwise or counterclockwise currents.
The phase boundary is calculated based on $J=0$. For small $N$, it is estimated from the solution of the equation: $2\alpha^{1-N}+k_{c}N-2\alpha =0$~\footnote{The equation is derived from $J=0$ in Eq.~(\ref{eq:Jtij}) and the explicit solution can be obtained using Lambert's W function}, while for large $N$, the boundary is estimated from Eq.~(\ref{eq:Jfp}) with $J=0$.
Both the estimated boundaries agree well with the simulation.
Figure~\ref{fig:pds} (c) and (d) clearly show that the phase boundary is independent of $u$ provided $T_{N-1}<1/Nu$ holds.

Here, we demonstrate that the discreteness in molecular number can induce a reversal of chemical current. 
In contrast to the small-number effect in the steady state distribution (e.g. emergence of bistability as shown in Fig.~\ref{fig:2stateTKs}-(b)), 
the reversal of chemical current is a characteristic small-number effect to a innately non-equilibrium system.
This phenomenon can emerge for a relatively large $N$, as is seen in Fig.~\ref{fig:TK3}-(d), which does not require an absolute small-number, such as $0, 1$ or $2$.
\section{2TK model with external flow}
\begin{figure}[h] 
\includegraphics[width=8cm]{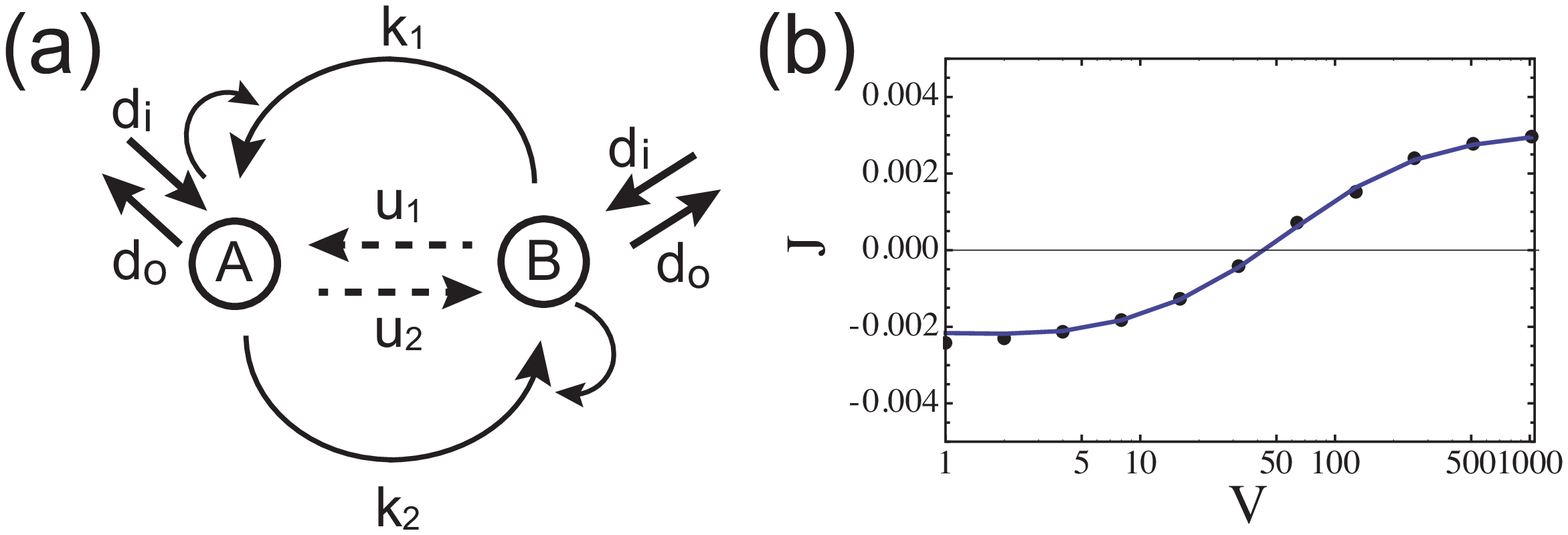}
\caption{Behaviors of the 2TK model under external flow. (a) Schematic representation of the 2TK model under external flow. 
(b) Dependence of current of the chemical reaction $B \to A$ on volume $V$. Parameters $k_{1} = 1, k_{2} = 0.99, u_{1} = 0.001, u_{2} = 0.005, d_{i} = 0.01$, and  $d_{o} = 0.01$ are used. The black points represent simulation results, while the thick line shows the analytical results. With these parameters, the chemical reaction current $J$ shows a reversal at approximately $V=40$.}
\label{fig:open2TK}
\end{figure}
Reversal of flow by smallness in molecule number in the last section is a novel discovery, but a set of  reactions involved therein may be a bit complicated to be realized. In this section, we show generality of such discreteness-induced reversal, by taking a two-component autocatalytic reaction system under external flow.  To be specific, we extend the 2TK model under external flow so that the total number of molecules is not conserved, to demonstrate both the applicability of our analytic method and generality of the reversal phenomena.
 As shown in Fig.~\ref{fig:open2TK}-(a), this model is composed of two chemicals, A and B, and the following 8 reactions: 
\begin{eqnarray}
&A+B \xrightarrow[k_{1}]{} 2A,& A+B \xrightarrow[k_{2}]{} 2B \\
&A \xrightarrow[u_{2}]{} B,& B \xrightarrow[u_{1}]{} A\nonumber \\
&A \xrightarrow[d_{o}]{} \phi,& \phi \xrightarrow[d_{i}]{} A\nonumber \\
&B \xrightarrow[d_{o}]{} \phi,& \phi \xrightarrow[d_{i}]{} B\nonumber.
\end{eqnarray}	
Here the annihilation and creation of $A$ and $B$ in the last two lines are added to the 2TK model.
The master equation of this model is given by
\begin{eqnarray}
\dot{P}_{n,m}&=& T^{n+}_{n-1}P_{n-1,m}-T^{n+}_{n}P_{n,m} \nonumber \\
&&+T^{n-}_{n+1}P_{n+1,m}-T^{n-}_{n}P_{n,m} \\
&&+T^{m+}_{m-1}P_{n,m-1}-T^{m+}_{m}P_{n,m} \nonumber \\
&&+T^{m-}_{m+1}P_{n,m+1}-T^{m-}_{m}P_{n,m} \nonumber \\
&&+T^{n+,m-}_{n-1,m+1}P_{n-1,m+1}-T^{n+,m-}_{n,m}P_{n,m} \nonumber \\
&&+T^{n-,m+}_{n+1,m-1}P_{n+1,m-1}-T^{n-,m+}_{n,m}P_{n,m}, \nonumber 
\end{eqnarray}
where the transition rates are given by
\begin{eqnarray} \label{eq:tp}
&T^{n+}_{n}=d_{i}V,&  T^{n-}_{n}=d_{o}n\\
&T^{m+}_{m}=d_{i}V,& T^{m-}_{m}=d_{o}m\nonumber \\
&T^{n+,m-}_{n,m}=u_{1}m+k_{1}\frac{nm}{V},& T^{n-,m+}_{n,m}=u_{2}n+k_{2}\frac{nm}{V}. \nonumber
\end{eqnarray}
In this model, $V$ represents the volume of the system. For infinitely large $V$, the model is described by the following rate equations:
\begin{eqnarray}
\dot{a} &=& (k_{1}-k_{2})ab+u_{1}b-u_{2}a+d_{i} -d_{o}a \\
\dot{b} &=& (k_{2}-k_{1})ab-u_{1}b+u_{2}a+d_{i} -d_{o}b.\nonumber
\end{eqnarray}
The major difference between the 2TK model with and without external flow is that the former model has a chemical reaction flow $J$ from A to B as calculated by $J=(k_{1}-k_{2})a^{*}b^{*}+u_{1}b^{*}-u_{2}^{*}a$, in which $a^{*}$ and $b^{*}$ represent the steady state solutions of the above equations, while in the 2TK model without external flow, $J$ is always zero.

Similar to the 3-component model that we proposed, this 2TK model under external flow shows significant deviation from the chemical reaction current estimated by the rate equation, when the system size $V$ becomes small.  
Specifically, a reversal of the chemical reaction flow is observed for small $V$, when $(k_{1}-k_{2})a^{*}b^{*}+u_{1}b^{*}-u_{2}a^{*}$ and $(u_{1}b^{*}-u_{2}a^{*})$ have different signs. Figure~\ref{fig:open2TK}-(b) shows an example of the reversal of $J$ against the change in $V$. 

The 2TK model under external flow satisfies neither the requirement of conservation of the total number of molecules nor cyclic symmetry, as is the case in the 3-component model. Therefore, it is difficult to describe the chemical reaction current as a successive fixation of one of the chemical species, as was accepted for the 3-component model analysis. Here, we apply an alternative approach to mapping from the two-variable Markov process to a one-variable Markov process. 
Recalling that the rate of change in transition probability following the application of $n\to n \pm1$ (e.g., $(T^{n+}_{n}-T^{n+}_{n\pm1})/T^{n+}_{n}$) is negligible when $n$ is large, we assume that $P_{n\pm 1,m} \sim P_{n,m}$ for $n>m$ and  $P_{n,m\pm 1} \sim P_{n,m}$ for $n  \le m$. We also assume that $d$, the reaction rate for $\phi \to A$ and $\phi \to B$, is sufficiently small. With these assumptions, the two-variable master equation is reduced to a one-variable master equation, such that
\begin{eqnarray}
\hspace{-0.5cm}
\dot{P}_{n,m} = \lambda_{n-1}P_{n-1,m+1}-\lambda_{n}P_{n,m}+\mu_{n+1}P_{n+1,m-1}-\mu_{n}P_{n,m}, \label{eq:master1}
\end{eqnarray}
where 
\begin{eqnarray}
  \begin{cases}
\lambda_{n}&=u_{1}m+k_{1}\frac{nm}{V}+d_{i}V\\
 \mu_{n,m}&=u_{2}n+k_{2}\frac{nm}{V}+d_{o}n 
 \end{cases}
\ \   \mbox{for} \ n\le m, \\
  \begin{cases}
  \lambda_{n}&=u_{1}m+k_{1}\frac{nm}{V}+d_{o}m \\
   \mu_{n,m}&=u_{2}n+k_{2}\frac{nm}{V}+d_{i}V 
 \end{cases}
 \ \ \ \mbox{for} \ n > m.
\end{eqnarray}
The  reduced model given in Eq. (\ref{eq:master1}) indicates that the total number of molecules is conserved as $N=n+m$ and, thus, $m$ can be expressed as $m=N-n$. By $\dot{P}_{n,m}=0$, the steady state distribution of $P_{n,m}$ with boundary condition $\mu_{1,N-1}P_{1,N-1}-\lambda_{0,N}P_{0,N}=0$ is obtained as
\begin{equation}
P_{n,m}= \frac{\prod_{j=1}^{n} \lambda_{j-1}/\mu_{j}}{\sum_{i=0}^{N} \prod_{j=1}^{i} \lambda_{j-1}/\mu_{j}} \rho(N),
\end{equation}
where $\rho (N)$ is the probability of $n+m=N$. 
Equation.(\ref{eq:tp}) indicates that the transition probability of $N \to N+1$ is $2d_{i}V$ and that of $N \to N-1$ is $d_{o}N$, indicating that $\rho (N)$ is the Poisson distribution $\rho (N)=e^{-\Lambda}\Lambda^{N}/N!$, where $\Lambda=2d_{i}V/d_{o}$.

The chemical reaction flow $J$ from A to B is calculated from 
\begin{eqnarray}
J&=&\sum_{N=0}^{\infty}\sum_{n=0}^{N}\left( u_{1}\frac{N-n}{V}+k_{1} \frac{n(N-n)}{V^{2}} \right. \nonumber \\ 
&& \left. - u_{2}\frac{n}{V}-k_{2} \frac{n(N-n)}{V^{2}} \right)P_{n,N-n}, \nonumber
\end{eqnarray}
and the estimated $J$ is shown in Fig.~\ref{fig:open2TK}-(b). Note, however, that this approximation is only valid for sufficiently small $d$. 

Here, we have shown that the change in chemical reaction flow due to the small-number effect generally occurs in the presence of the reaction $A+B \to 2A/2B$. 
For some parameter values, the current of chemical reaction $B \to A$ is even reversed with the decrease in molecule number.

\section{Discussion}
In this paper, we have investigated small-number effects in chemical reactions using a method based on a single-variable Markov process.  
After analyzing the 2TK model, we have analyzed the 3-component model and a two component model under external flow in order to demonstrate a novel type of small-number effect, i.e.,
the reversal of the reaction current. 
This reversal is first reported here and demonstrates a small-number effect in a non-equilibrium system.
The 3-component model here consists of homogeneous cyclic reactions, but extensions to chain-like reactions and systems with inhomogeneous reaction coefficients~\cite{togashi2003alteration} are straightforward. In the case of chain-like reactions, the dominant molecule in the steady state may change depending on $N$ because the direction of the reaction current may change.  
Although the proposed 3-component model includes relatively complex chemical reactions, the reversal in reaction current is not specific to this model.
Indeed, in Sec.IV, we have shown that a simple autocatalytic reaction set can also exhibit the reversal of chemical current.

The emergence of multi-stability induced by the small-number effect in a real biological system has been found and has attracted much interest~\cite{artyomov2007purely, altschuler2008spontaneous,ma2012small,to2010noise}, with most of the chemical reaction systems examined in those studies having autocatalytic reactions.
As autocatalytic reactions can potentially show the reversal in chemical current,
the reversal (or, at least, change) in chemical current we proposed here will be observed in these biological systems also, and provides a novel mechanism of controlling the reaction process based on the number of molecules within the system, rather than their concentration. 
\section*{ACKNOWLEDGMENT}
This work was supported by a Grant-in-Aid for Scientific Research on Innovative Areas “Spying minority in biological phenomena (No.3306)” (26115704), and "the Platform for Dynamic Approaches to Living System" from The Ministry of Education, Culture, Sports, Science, and Technology, Japan. We acknowledge helpful discussions with Shuji Ishihara.
\bibliographystyle{apsrev4-1}
%

\end{document}